\documentclass[preprint,showpacs,amsmath,amssymb,prd,nofootinbib]{revtex4}
\usepackage{epsf}
\usepackage{dcolumn}
\usepackage{amsmath}
\usepackage{amssymb}
\usepackage{graphicx}

\usepackage[breaklinks=true,colorlinks=true]{hyperref}

\begin{document}

\title{On the ``old'' and ``new'' relativistic wave equations for the particle having spin s=3/2}

\author{Volodimir Simulik $^{1,2}$, Ilona Vyikon $^{1}$}

\email{vsimulik@gmail.com}
\affiliation{$^{1}$  Institute of Electron Physics, National Academy of Sciences of Ukraine, 21 Universitetska Str. 88017 Uzhgorod, Ukraine}
\affiliation{$^{2}$ Erwin Shr\"odinger International Institute for Mathematics and Physics, University of Vienna}
\date{May 2022}


\begin{abstract}
Relativistic wave equation of motion without redundant components for the particle having spin $3/2$ has been considered. In order to show the newness a comparison with the known equations for the spin $s$=3/2 field is given. Therefore, the brief review of the relativistic wave equations for the particle with spin $s$=3/2 is suggested. In our equation the wave function for the particle-antiparticle doublet contains only 8 components. The consideration is carried out both at the level of relativistic quantum mechanics and at the level of local field theory. The extended Foldy--Wouthuysen transformation, which gives the operator link between these two levels is suggested.    
\end{abstract}

\pacs {03.65.Pm, 11.30.-z, 11.30.j, 04.20.Cv}

\maketitle

\section{Introduction}

We consider the first order relativistic equations [1--12], which describe the motion of the elementary particle having spin s=3/2. A comparative characteristic of such equations is given. The second order equations of the Klein--Gordon type are not discussed.

Known first order relativistic equations of motion of an elementary particle with spin s=3/2 [1--9] can be divided into two groups. The first group is a partial case of the corresponding equations for an arbitrary spin. The second group includes the specially suggested equations for the description of the properties of an elementary particle having spin s=3/2.

Recently, in order to describe the motion of a particle of arbitrary spin better, the start of introducing of the new equation [10--12] was given. Such equation has an interesting special case for the spin s=3/2. The first step in analysis of the equation [10--12] is the comparison with the known approaches to the problem. Below, in order to provide the comparison of the main properties of the equations [1--12] between each other, the partial case when s=3/2 is chosen.

The first equation for a particle of an arbitrary half-integer spin was proposed by P. Dirac in 1936 [1]. After that step by step main equations for an arbitrary spin by M. Fierz and W. Pauli [2,3], by H. Bhabha [4--6] and by W. Bargmann and E. Wigner [7] were suggested.

In the first group the Bhabha [4--6] and Bargmann--Wigner [7] equations are the most known. The Rarita--Schwinger [8] and Fisk--Tait [9] equations are the most known among the special equations (second group) for the particle having spin 3/2.

Note that hadrons with spin 3/2 are long known in the particle physics. In particular, there is a multiplet of $\Delta$-baryons, or $\Delta$-resonances ($\Delta^{++}, \Delta^{+},\Delta^{0},\Delta^{-}$), consisting of three quarks of the type u and d. The $\Omega^{-}$ hyperon consists of three strange quarks (sss) and has a lifetime $10^{-10}$s. Effective quantum field theory describes some properties of such elementary particles on the basis of the Rarita--Schwinger equation [13--15] and its minor modifications [16]. A fundamentally different physical example arises in supersymmetry. It is gravitino, which is a superpartner of graviton. However, contrary to hyperons (baryons with the spin 3/2), superpartners are not experimentally observable. Moreover, the graviton itself generates much more questions than the clear answers exist.

Registration of $\Delta$-resonances in LHC experiments continues. So we have new candidates [17, 18] for the particles with spin 3/2. Furthermore, elementary particles with s=3/2 are often considered [19--21] as a carriers of dark matter. Finally, spin 3/2 is found not only in elementary particles but also in other physical systems [22, 23].

It should be noted especially that the theoretical description of the elementary particles with spin 3/2 in the frameworks of field theory still deals with a number of fundamental problems (see, e.g., refs. [24--31], as well as [13--16]). This is typical for all fields with higher spins that are larger than s=1. The main difficulties are caused by the fact that, in addition to the components related to the spin s=3/2, the corresponding quantum field also has redundant components. Moreover, in equations [1--6, 8,9]  the representations of the Lorenz group, not of the Poincar{\'e} group, are applied. However, the Poincar{\'e} symmetry of the Rarita--Schwinger system of equations is proved as well, see, e.g., refs. [32, 33].

\section{The Pauli--Fierz equation}

In [3] both the field of arbitrary spin and the partial cases, when the values of spin are 3/2 or 2, were considered. Tensors are used to describe the case of an integer spin, and spinors are used to describe a half-integer spin. W. Pauli and M. Fierz started from the Dirac equation [1]. Therefore, sometimes this object is called as the Dirac--Pauli--Fierz equation.

In the absence of external forces, the wave field corresponding to the particles with spin 3/2 is described by the following spinors (below we use the notation from [3])
\begin{equation}
\label{Eg.1}
a^{\dot{\alpha}}_{\beta\gamma}=a^{\dot{\alpha}}_{\gamma\beta}, \quad b^{\dot{\alpha}\dot{\beta}}_{\gamma}=b^{\dot{\beta}\dot{\alpha}}_{\gamma}.
\end{equation}
Spinors (1) go into one another by reflections and satisfy the equations 
\begin{equation}
\label{Eg.2}
p^{\dot{\beta}\rho}a^{\dot{\alpha}}_{\rho\gamma}+p^{\dot{\alpha}\rho}a^{\dot{\beta}}_{\rho\gamma}=2mb^{\dot{\alpha}\beta}_{\gamma}, \quad p_{\alpha\dot{\rho}}b^{\dot{\rho}\dot{\gamma}}_{\beta}+p_{\beta\dot{\rho}}b^{\dot{\rho}\dot{\gamma}}_{\alpha}=2ma^{\dot{\gamma}}_{\alpha\beta},
\end{equation}
together with the conditions
\begin{equation}
\label{Eg.3}
p_{\dot{\alpha}}\mbox{}^{\beta}a^{\dot{\alpha}}_{\beta\gamma}=0, \quad p_{\dot{\alpha}}\mbox{}^{\gamma}b^{\dot{\alpha}\dot{\beta}}_{\gamma}=0.
\end{equation}
Here the operator of momentum has the specific form $p_{\dot{\alpha}\beta}=-i\sigma_{\dot{\alpha}\beta}^{k}\frac{\partial}{\partial x_{k}}$, and $\sigma_{\dot{\alpha}\beta}^{k}=\left(\sigma^{1},\sigma^{2},\sigma^{3},iI\right)_{\dot{\alpha}\beta}$ are the Pauli matrices in standard representation. Matrix $\sigma_{k}^{\dot{\alpha}\beta}$ is the Hermitian conjugate of $-\sigma_{\alpha\dot{\beta}}^{k}$. The second order wave equation for $a^{\dot{\gamma}}_{\alpha\beta}$ and $b^{\dot{\alpha}\dot{\beta}}_{\gamma}$ follows from these equations. The additional conditions (3) mean that formalism is without particles having spin 1/2. 

Equations (1) and (2) were derived in [3] from the variation principle as well. Corresponding Lagrange formalism led to the differential relationship between the spinors $a^{\dot{\gamma}}_{\alpha\beta}$ and $b^{\dot{\alpha}\dot{\beta}}_{\gamma}$ in the form of equations as follows: $p^{\dot{\beta}\rho}a^{\dot{\alpha}}_{\gamma\rho}=mb^{\dot{\alpha}\dot{\beta}}_{\gamma}, \, p_{\alpha\dot{\rho}}b^{\dot{\rho}\dot{\gamma}}_{\beta}=ma^{\dot{\gamma}}_{\alpha\beta}$.

The wave function in the system of equations (2) and (3) is a matrix-column with 16 rows. The Pauli--Fierz equation for the spin case 3/2 is equivalent to the Rarita--Schwinger equation, which we consider below. Such equivalence has been noted by a number of authors, see, for example, [34] and a review in [33].

Note that in [35] the Rarita--Schwinger equation was derived directly from the Pauli--Fierz equation. Unfortunately, the author of [35] forgot to refer to the article of Rarita--Schwinger [8], which at that time was already well known.

\section{The Bhabha equation}

This equation is known from [4], see also [5]. Such matrix-differential equation in partial derivatives of the first order has a general form for arbitrary spin. Nevertheless, for each partial value of the spin the main matrices have different explicit forms, which depend on the different explicit types of generators of the Lie algebra of the corresponding representation of the Lorentz group.

The Bhabha equation has the form
\begin{equation}
\label{Eg.4}
\left(p_{\mu}\alpha^{\mu}+m\right)\psi\left(x\right)=0, \quad \mu=\overline{0,3} \quad j=1,2,3,
\end{equation}
where $p_{\mu}\equiv i\partial/\partial x^{\mu}$, $m$ is an arbitrary constant (the mass of the particle, and different mass values for particles of different spins are possible), $\alpha^{\mu}$ are four matrices that satisfy different sets of commutation relations in each case. Such equation is invariant with respect to arbitrary transformations of the Lorentz group if the matrices $\alpha^{\mu}$ obey the commutation relations
\begin{equation}
\label{Eg.5}
\left[\alpha^{\mu},S^{\rho\sigma}\right]=\alpha^{\mu}S^{\rho\sigma}-S^{\rho\sigma}\alpha^{\mu}=g^{\mu\rho}\alpha^{\sigma}-g^{\mu\sigma}\alpha^{\rho},
\end{equation}
where metric tensor $g^{\mu\nu}$ is defined as $g^{00}=-g^{11}=-g^{22}=-g^{33}=$I, and matrices $S^{\rho\sigma}=-S^{\sigma\rho}$ determine six generators of transformations of a concrete representation of the Lie algebra of the Lorentz group, which satisfy the commutation relations in the form
\begin{equation}
\label{Eg.6}
\left[S^{\mu\nu},S^{\rho\sigma}\right]=-g^{\mu\rho}S^{\nu\sigma}+g^{\mu\sigma}S^{\nu\rho}+g^{\nu\rho}S^{\mu\sigma}-g^{\nu\sigma}S^{\mu\rho}.
\end{equation}
The following condition $\left[\alpha^{\mu},\alpha^{\nu}\right]=S^{\mu\nu}$ is necessary as well. 

The number of components of the wave function in equation (4) is different for different values of spin. In [6] Bhabha described the partial case, when $s$ = 3/2, and personally proved that in this case his equation coincides with the Rarita--Schwinger equation. In particular, the wave function has 16 components. Note also that Bhabha and his followers used the representations of the Lorentz group, not the representations of the Poincar{\'e} group.

\section{The Bargmann--Wigner equation}

The equation from [7], as it is evident. e.g., from [36], is a Dirac-like equation in spaces of arbitrary dimensions, in which the gamma matrix representations of Clifford algebra over the field of complex numbers are determined.

The Bargmann--Wigner wave function for a particle of arbitrary spin s is a multispinor $\psi_{\alpha_{1}\alpha_{2}...\alpha_{2s}}$ with 2s spinor indices, each running independently from 1 to 4. Such wave function is completely symmetrical under permutations of indices. The wave equation is given by
\begin{equation}
\label{Eg.7}
\left(\gamma^{(n)}_{\mu}p^{\mu}-m\right)\psi\left(x\right)=0, \quad  n=1,...,2s,
\end{equation}
where the $\gamma^{(n)}_{\mu}$ are the set of Dirac $\gamma$-matrices operating on $n$-th spinor index of $\psi$. More precisely, if $\psi$ is regarded as a vector in the direct product space of $2s\cdot 4$-dimensional spaces (corresponding to the $2s$ spinor indices), and if we denote by I and $\gamma_{\mu}$ the unit matrix and the usual Dirac matrices in any one of these factor spaces, then $\gamma^{(n)}_{\mu}$ is the Kronecker product $\gamma^{(n)}_{\mu}=\mathrm{I}\times ... \times \gamma_{\mu}\times ... \times \mathrm{I}$, with $\gamma_{\mu}$  occurring as the $n$-th factor. The dimension of such spaces is $2s\times4$, hence, for $s$=3/2 the wave function will have the 12 components.

Note a special development of the Bargmann--Wigner formalism for the case of a free spin-3/2-particle, which was published in [37]. A comparison with the Rarita--Schwinger theory is discussed. It is shown that the theories are equivalent. 

\section{The Rarita--Schwinger equation}

This equation describes the particle with arbitrary half-integer spin but the most known is the application to the particle having spin 3/2.

Thus, in [8], a fundamentally new object was introduced, which has four spinor indices and additional indices that define a symmetric tensor of arbitrary rank k. It is the object that acts as a wave function in the equation from [8]. This is the combination of two different elements of the Pauli--Fierz formalism [3] in order to simplify it. For k=0 one has ordinary Dirac equation, in the case, when k=1, it gives the 16-component equation for the particle having spin 3/2. Such equation has some relation to the description of fermions with spin 3/2 only if the additional conditions are imposed. Namely this equation is widespread and is called the Rarita--Schwinger equation [8] for a particle with spin 3/2. In addition to the existence of redundant components, problems naturally arise with the transformational properties of the 16-component vector-spinor, causality, relativistic invariance and quantization. The main problem is an interaction with external fields. These difficulties of the Rarita--Schwinger equation were mentioned by many researchers, see, for example, [24--31] . 

Hence, the Rarita--Schwinger equation without any loss of generality can be written as
\begin{equation}
\label{Eg.8}
\left(i\gamma^{\mu}\partial_{\mu}-m\right)\psi_{\nu}-\frac{1}{3}\left(\gamma_{\nu}i\partial_{\mu}+\gamma_{\mu}i\partial_{\nu}\right)\psi^{\mu}+\frac{1}{3}\gamma_{\nu}\left(i\gamma^{\alpha}\partial_{\alpha}+m\right)\left(\gamma^{\mu}\psi_{\mu}\right)=0.
\end{equation}
After imposing additional conditions the system of equations is obtained as follows
\begin{equation}
\label{Eg.9}
\left(i\gamma^{\mu}\partial_{\mu}-m\right)\psi_{\nu}\left(x\right)=0, \quad  \partial^{\mu}\psi_{\mu}=0, \quad \gamma^{\mu}\psi_{\mu}=0.
\end{equation}

\section{The Fisk--Tait equation}

This equation was suggested in [9] for the particle having spin 3/2. It is the Dirac-like equation with 24-component wave function in the form of antisymmetric tensor-spinor of the Lorentz group
\begin{equation}
\label{Eg.10}
\Psi^{\mu\nu}=\left( 
\begin{array}{cccc}
0 & \Psi^{01} & \Psi^{02} & \Psi^{03} \\ 
-\Psi^{01} & 0 & \Psi^{12} & \Psi^{13} \\ 
-\Psi^{02} & -\Psi^{12} & 0 & \Psi^{23} \\ 
-\Psi^{03} & -\Psi^{13} & -\Psi^{23} & 0
\end{array}
\right).
\end{equation}
Here each component of the tensor (10) is the 4-component Dirac spinor. In notations
\begin{equation}
\label{Eg.11}
\psi=\left(\Psi^{01},\Psi^{02},\Psi^{03}\right) \quad \chi=\left(\Psi^{23},\Psi^{31},\Psi^{12}\right),
\end{equation}
the Fisk--Tait equation has the form
\begin{equation}
\label{Eg.12}
\left(\gamma^{\rho}p_{\rho}-m\right)\psi^{\mu\nu}\left(x\right)=0, \quad \gamma_{\mu}\gamma_{\nu}\psi^{\mu\nu}=0, \quad \varepsilon^{\mu\nu}\mbox{}\mbox{}_{\sigma\rho}p_{\nu}\psi^{\sigma\rho}=0,
\end{equation}
$$\left(\gamma^{\rho}p_{\rho}-m\right)\chi^{\mu\nu}\left(x\right)=0, \quad \gamma_{\mu}\gamma_{\nu}\chi^{\mu\nu}=0, \quad p_{\mu}\chi^{\mu\nu}=0.$$

The system of equations (12) was put into consideration in order to overcome the difficulties mentioned in [24--31]. Note only partial and rather conditional success. The equation is criticized in [38] for doubling the parity and the presence of negative energy. Nevertheless, this approach still has followers up for today [39].

Improvement of the models [3, 5, 7--9]  considered here is still relevant, see, e.g. [16, 39].

\section{Equation without redundant components}

The relativistic quantum-mechanical equation [10, 11] for the spin s=3/2 fermion-antifermion doublet is given by
\begin{equation}
\label{Eg.13}
i\partial_{0}f\left(x\right)=\sqrt{-\Delta +m^{2}}f\left(x\right), \quad f=\mathrm{column}\left|f^{1},f^{2},f^{3},f^{4},f^{5},f^{6},f^{7},f^{8}\right|.
\end{equation}
The general solution has the form
\begin{equation}
\label{Eg.14}
f(x)= \left|
{{\begin{array}{*{20}c}
 f_{\mathrm{part}} \hfill  \\
 f_{\mathrm{antipart}} \hfill  \\
\end{array} }} \right| =\frac{1}{\left(2\pi\right)^{\frac{3}{2}}}\int d^{3}k e^{-ikx}b^{\mathrm{A}}(\overrightarrow{k})\mathrm{d}_{\mathrm{A}}, \quad \mathrm{A}=\overline{1,8},
\end{equation}
where $\mathrm{d}_{\mathrm{A}}$ are the orts of 8-component Cartesian basis: $\mathrm{d}_{\mathrm{A}}=\left\{\delta_\mathrm{A\dot{B}}\right\},\, \mathrm{A\dot{B}}=\overline{1,8}$. The functions $b^{1}(\overrightarrow{k}), \,
b^{2}(\overrightarrow{k}), \, b^{3}(\overrightarrow{k}), \, b^{4}(\overrightarrow{k})$ in
solution (14) are the momentum-spin amplitudes of the massive
fermion with the spin s=3/2 and the spin projection $(3/2,1/2,-1/2,-3/2)$,
respectively; $b^{5}(\overrightarrow{k}), \,
b^{6}(\overrightarrow{k}),$ $b^{7}(\overrightarrow{k}), \, b^{8}(\overrightarrow{k})$ are the
momentum-spin amplitudes of the antiparticle (antifermion) with the
spin s=3/2 and the spin projection $(-3/2,-1/2,1/2,3/2)$, respectively. Such interpretation of the amplitudes directly follows from the equations on eigenvalues of the momentum and spin operators (the explicit form of the spin operator is given just below in the formulas (15)).

Equation (13) is considered in rigged Hilbert space $\mathrm{S}^{3,8}\subset\mathrm{H}^{3,8}\subset\mathrm{S}^{3,8*}$, where $\mathrm{H}^{3,8}$ is the Hilbert space of 8-component functions, $\mathrm{S}^{3,8}$ is the corresponding space of Schwartz test functions, which is dense in Schwartz generalized function space $\mathrm{S}^{3,8*}$ (space $\mathrm{S}^{3,8*}$ is conjugated to $\mathrm{S}^{3,8}$ by the corresponding topology).

We call the model of the physical reality, which is based on the equation (13), as the relativistic canonical quantum mechanics of the spin s=3/2 fermion-antifermion doublet. Indeed, equation (13) directly demonstrates the relativistic relationship between energy, momentum and the mass of the particle, does not lead to negative energies and describes SU(2) spin of antiparticle as a mirror reflection of particle SU(2) spin:
\begin{equation}
\label{Eg.15}
s^{1}_{8}=\left( 
\begin{array}{cccc}
s^{1} & 0  \\ 
0 & -s^{1} \\ 
\end{array}
\right), \quad s^{2}_{8}=\left( 
\begin{array}{cccc}
s^{2} & 0  \\ 
0 & s^{2} \\ 
\end{array}
\right), \quad s^{3}_{8}=\left( 
\begin{array}{cccc}
s^{3} & 0  \\ 
0 & -s^{3} \\ 
\end{array}
\right),
\end{equation}
where
\begin{equation}
\label{Eg.16}
s^{1}=\frac{1}{2}\left( 
\begin{array}{cccc}
0 & \sqrt{3} & 0 & 0 \\ 
\sqrt{3} & 0 & 2 & 0 \\ 
0 & 2 & 0 & \sqrt{3} \\ 
0 & 0 & \sqrt{3} & 0
\end{array}
\right), \, s^{2}=\frac{i}{2}\left( 
\begin{array}{cccc}
0 & -\sqrt{3} & 0 & 0 \\ 
\sqrt{3} & 0 & -2 & 0 \\ 
0 & 2 & 0 & -\sqrt{3} \\ 
0 & 0 & \sqrt{3} & 0
\end{array}
\right), \, s^{3}=\frac{1}{2}\left( 
\begin{array}{cccc}
3 & 0 & 0 & 0 \\ 
0 & 1 & 0 & 0 \\ 
0 & 0 & -1 & 0 \\ 
0 & 0 & 0 & -3
\end{array}
\right). 
\end{equation}
Here antiparticle is related to four lowest components of the column from (13).

It is easy to verify that the commutation relations $\left[s^{j}_{8},s^{\ell}_{8}\right]=i\varepsilon^{j \ell n}s^{n}_{8}$ of the SU(2)-algebra are valid. The Casimir operator has the form of the following $8 \times 8$ diagonal matrix $\overrightarrow{s}^{2}_{8}= \frac{15}{4}
\mathrm{I}_{8}=\frac{3}{2}\left(\frac{3}{2}+1\right)\mathrm{I}_{8}$, where $\mathrm{I}_{8}$ is the $8\times 8$ unit matrix.

We put into consideration, for the particle having spin s=3/2, the equation that is an analogue of the Dirac equation for the fermion with spin 1/2 and nonzero mass. In the form of Hamilton this equation is given by
\begin{equation}
\label{Eg.17}
\left[i\partial_{0}-\Gamma_{8}^{0}(\overrightarrow{\Gamma}_{8}\cdot \overrightarrow{p}+m)\right]\psi(x)=0, \quad \psi=\mathrm{column}\left|\psi^{1},\psi^{2},\psi^{3},\psi^{4},\psi^{5},\psi^{6},\psi^{7},\psi^{8}\right|.
\end{equation}
where the $8\times 8$ gamma matrices have the form
\begin{equation}
\label{Eg.18}
 \Gamma_{8}^{0}=\left| {{\begin{array}{*{20}c}
 \mathrm{I}_{4} \hfill & 0 \\
 0 \hfill & -\mathrm{I}_{4} \\
\end{array} }} \right|, \quad \Gamma_{8}^{j}=\left| {{\begin{array}{*{20}c}
 0 \hfill & \Sigma^{j} \\
 -\Sigma^{j} \hfill & 0 \\
\end{array} }} \right|,
\end{equation}
with
\begin{equation}
\label{Eg.19}
\Sigma^{j}=\left| {{\begin{array}{*{20}c}
 \sigma^{j} \hfill & 0 \\
 0 \hfill & \sigma^{j} \\
\end{array} }} \right|, 
\end{equation}
and $\sigma^{j}$ are the standard $2\times 2$ Pauli matrices.

The equation (17) is derived from the relativistic canonical quantum mechanics (13) on the basis of transformation, which is given by the extended Foldy--Wouthuysen operator $V$ in the form:
\begin{equation}
\label{Eg.20}
V\left(\partial_{0}+i\omega\right)V^{-1}=\partial_{0}+i\Gamma_{8}^{0}(\overrightarrow{\Gamma}_{8}\cdot \overrightarrow{p}+m), \quad \psi = Vf; \quad \omega=\sqrt{-\Delta +m^{2}}=\sqrt{\overrightarrow{p}^{2}+m^{2}}.
\end{equation}
Operator $V$ is given by
\begin{equation}
\label{Eg.21}
V=\frac{
i\Gamma^{j}_{8}\partial_{j}+\omega+m}{\sqrt{2\omega(\omega+m)}}\left| {{\begin{array}{*{20}c}
 \mathrm{I}_{4} \hfill & 0 \\
 0 \hfill & \mathrm{I}_{4}C \\
\end{array} }} \right|, \quad V^{-1}=\left| {{\begin{array}{*{20}c}
 \mathrm{I}_{4} \hfill & 0 \\
 0 \hfill & \mathrm{I}_{4}C \\
\end{array} }} \right|\frac{
-i\Gamma^{j}_{8}\partial_{j}+\omega+m}{\sqrt{2\omega(\omega+m)}},
\quad VV^{-1}=V^{-1}V=\mathrm{I}_{8}, 
\end{equation}
where $\mathrm{I}_{4}$ is $4 \times 4$ unit matrix and $C$ is the operator of complex conjugation, $C \psi = \psi^{*}$ (operator of involution in the space $\mathrm{H}^{3,1}$). The transformation inverse to (20), (21) is valid as well.

Note that transformation (20), (21) can be applied only to the operators (of equation, energy, momentum, spin, etc.) taken in anti-Hermitian form. 

The general solution of the equation (17) is given by
\begin{equation}
\label{Eg.22}
\psi(x)=\frac{1}{\left(2\pi\right)^{\frac{3}{2}}}\int d^{3}k\left[e^{-ikx}c^{\mathrm{A}}(\overrightarrow{k})\mathrm{v}^{-}_{\mathrm{A}}(\overrightarrow{k})+e^{ikx}c^{*\mathrm{B}}(\overrightarrow{k})\mathrm{v}^{+}_{\mathrm{B}}(\overrightarrow{k})\right],
\end{equation}
where $\mathrm{A}=\overline{1,4}, \, \mathrm{B}=\overline{5,8}$ and the 8-component spinors $(\mathrm{v}^{-}_{\mathrm{A}}(\overrightarrow{k}), \, \mathrm{v}^{+}_{\mathrm{B}}(\overrightarrow{k}))$ are given by
\small
$$\mathrm{v}^{-}_{1}(\overrightarrow{k}) = N\left|
\begin{array}{cccc}
 \widetilde{\omega}+m \\
 0 \\
 0 \\
 0 \\
 k^{3} \\
 k^{1}+ik^{2} \\
 0 \\
 0 \\
\end{array} \right|, \quad \mathrm{v}^{-}_{2}(\overrightarrow{k}) = N\left|
\begin{array}{cccc}
 0 \\
 \widetilde{\omega}+m \\
 0 \\
 0 \\
 k^{1}-ik^{2} \\
 -k^{3} \\
 0 \\
 0 \\
\end{array} \right|,$$
$$ \mathrm{v}^{-}_{3}(\overrightarrow{k}) = N \left|
\begin{array}{cccc}
 0 \\
 0 \\
 \widetilde{\omega}+m \\
 0 \\
 0 \\
 0 \\
 k^{3} \\
 k^{1}+ik^{2} \\
\end{array} \right|,
\quad \mathrm{v}^{-}_{4}(\overrightarrow{k}) = N\left|
\begin{array}{cccc}
 0 \\
 0 \\
 0 \\
 \widetilde{\omega}+m \\
 0 \\
 0 \\
 k^{1}-ik^{2} \\
 -k^{3} \\
\end{array} \right|,$$
\begin{equation}
\label{Eg.23}
\mathrm{v}^{+}_{5}(\overrightarrow{k}) = N\left|
\begin{array}{cccc}
 k^{3} \\
 k^{1}+ik^{2} \\
 0 \\
 0 \\
 \widetilde{\omega}+m \\
 0 \\
 0 \\
 0 \\
\end{array} \right|, \quad
\mathrm{v}^{+}_{6}(\overrightarrow{k}) = N\left|
\begin{array}{cccc}
 k^{1}-ik^{2} \\
 -k^{3} \\
 0 \\
 0 \\
 0 \\
 \widetilde{\omega}+m \\
 0 \\
 0 \\
\end{array} \right|,
\end{equation}
$$\mathrm{v}^{+}_{7}(\overrightarrow{k}) = N\left|
\begin{array}{cccc}
 0 \\
 0 \\
 k^{3} \\
 k^{1}+ik^{2} \\
 0 \\
 0 \\
\widetilde{\omega}+m \\
 0 \\
\end{array} \right|, \quad
\mathrm{v}^{+}_{8}(\overrightarrow{k}) = N\left|
\begin{array}{cccc}
 0 \\
 0 \\
 k^{1}-ik^{2} \\
 -k^{3} \\
 0 \\
 0 \\
 0 \\
\widetilde{\omega}+m \\
\end{array} \right|,$$
\normalsize
where
\begin{equation}
\label{Eq.24} N\equiv
\frac{1}{\sqrt{2\widetilde{\omega}(\widetilde{\omega}+m)}}, \quad
\widetilde{\omega}\equiv \sqrt{\overrightarrow{k}^{2}+m^{2}}.
\end{equation}

The 8-component spinors (23) are derived from the orts
of the Cartesian basis with the help of the transformation (20), (21). The spinors (23) satisfy the relations of the orthonormalization and completeness similar to the corresponding relations for the standard 4-component Dirac spinors.

Function $c^{\mathrm{A}}(\overrightarrow{k})$, $c^{*\mathrm{B}}(\overrightarrow{k})$ in (22) are the momentum-spin amplitudes.  Quantum-mechanical interpretation of corresponding amplitudes can be given only in the frameworks of relativistic canonical quantum mechanics based on the equation (13).

It is easy to see that equation (17), contrary to other equations considered here, does not contain redundant components.

\section{Brief conclusions}

A comparison of the proposed new equation with known approaches to the description of a spin-3/2 particle is presented. Equation (17), whose wave function contains 8 components, has some advantages in comparison with other models discussed above, where wave functions have 12, 16 and 24 components. In addition to the absence of redundant components, the obvious advantage is the direct operator link (20), (21) to relativistic canonical quantum mechanics, which allows a clear quantum-mechanical interpretation of all statements, results and consequences.

Equation (13) itself has an independent meaning. Indeed, it is new equation for the particle having spin s=3/2 as well.

Briefly note the Poincar{\'e} invariance of equations (13), (17). If for equation (13) the corresponding representation of the Poincar{\'e} group is relatively simple, then for the Dirac-like equation (17) we have a somewhat cumbersome form (see [12], Chapter 7) of operators that define the Poincar{\'e} symmetry. Poincar{\'e}, not Lorentz, symmetry of equation (17) is its next advantage (the only exceptions are the Bargmann--Wigner equation and some special considerations of the Rarita--Schwinger equation), so an important task for further research is to simplify the explicit form of symmetry operators found in [12].

The proof of Poincar{\'e} invariance of equation (17) in a simple explicitly covariant form (not more cumbersome than for the usual Dirac equation, when we have 4-component spinors) is performed on the basis of the 256-dimensional representations [40] of the Clifford algebras $\textit{C}\ell^{\mathbb{R}}$(0,8), $\textit{C}\ell^{\mathbb{R}}$(1,7) (in the terms of $8 \times 8$  Dirac $\gamma$ matrices). On this basis we can prove not only spin 1 properties of the corresponding Dirac-like equation (17) but the spin 3/2 Poincar{\'e} symmetries as well. The method is known from [41--44]. The detailed consideration of the Poincar{\'e} invariance will be the task of the upcoming paper.

We have the reason to predict interesting and unexpected applications in high-energy physics and nuclear physics, especially for the problem of particle-antiparticle asymmetry at the beginning of the evolution of the Universe after a Big Bang. Indeed, equation (17) describes spin s=3/2 fermion-antifermion doublet.

\section{Acknowledgements}

Volodimir Simulik is very grateful for the two month Fellowship at the Erwin Shr\"odinger International Institute for Mathematics and Physics.




\end{document}